\documentclass{article}

 \usepackage[preprint]{neurips_2025}

\usepackage[utf8]{inputenc} 
\usepackage[T1]{fontenc}    
\usepackage{hyperref}       
\usepackage{url}            
\usepackage{booktabs}       
\usepackage{amsfonts}       
\usepackage{nicefrac}       
\usepackage{microtype}      
\usepackage{xcolor}         
\usepackage[pdftex]{graphicx}

\title{Prospects of Imitating Trading Agents\\in the Stock Market}

\author{%
  Mateusz Wilinski \\
  Department of Computing Sciences \\
  Tampere University \\ 
  Finland \\
  \texttt{mateusz.wilinski@tuni.fi} \\
  \And
  Juho Kanniainen \\
  Department of Computing Sciences \\
  Tampere University \\ 
  Finland \\
  \texttt{juho.kanniainen@tuni.fi}
}

\begin{document}

\maketitle

\begin{abstract}
In this work we show how generative tools, which were successfully applied to limit order book data, can be utilized for the task of imitating trading agents.
To this end, we propose a modified generative architecture based on the state-space model, and apply it to limit order book data with identified investors.
The model is trained on synthetic data, generated from a heterogeneous agent-based model.
Finally, we compare model's predicted distribution over different aspects of investors' actions, with the ground truths known from the agent-based model.
\end{abstract}

\section{Introduction}

Financial markets are complex adaptive systems \cite{hommes2001financial}, and their dynamics is a result of actions from many individuals.
These individuals have their own personal beliefs, goals, and constraints.
Moreover, they can interact with each other, adding to the complexity of the problem.
This pose a great challenge when trying to model such systems.
One approach to this problem is the agent-based modeling, where market participants are greatly simplified, but at the same time, a certain level of heterogeneity and individualism is possible \cite{helbing2012agent}.

Past agent-based models simplified behavioral patterns of individuals, due to limited computing power, limited analytical capabilities, many hidden variables, and/or lack of detailed knowledge.
However, with the the growing computing power and new machine learning capabilities, researchers start building more realistic models based on data \cite{pangallo2024data}.
The idea of calibrating agent-based models of stock market, using real data, is not new and was already present in the literature \cite{mike2008empirical}.
The main issue is the limited analytical tractability of such models.
In some cases, agent-based model can be translated into a probabilistic model with tractable likelihood \cite{monti2023learning}.
A more general approach is to use modern machine learning tools in order to approximate the posterior density and use it for inference \cite{dyer2022calibrating}.
Another way of combining agent-based modeling and machine learning in the stock market modeling business, is to create fast limit order book simulators \cite{byrd2020abides, belcak2021fast} in order to create synthetic environments for reinforcement learning tasks \cite{frey2023jax}.

Instead of following the bottom-up approach, like agent-based modeling, one can also directly apply machine learning machinery to replicate stock market behavior.
One promising direction is to use generative artificial intelligence models to predict the dynamics of the limit order book \cite{hultin2023generative, nagy2023generative, wheeler2024marketgpt}.
In this work we propose to apply similar generative models, but instead of predicting limit order book events, we try to predict individual investor future action.
This way we aim at building a generative model imitating investor behavior.
Such model can be further used to extend an existing agent-based model, and/or to gain a better understanding of individual investor strategy.
Following the idea from \cite{wilinski2025classifying}, we start by testing the prospects of imitating trading agents, by applying the generative approach to synthetic limit order book data.

\section{Methods}

\subsection{Synthetic data}\label{sec:data}

The data used for both training and evaluation is generated using the synthetic environment first introduced in \cite{wilinski2025classifying}.
The setting is the same as in the mentioned article, which means there are 1590 agents divided into 15 groups.
Each group has different trading strategy and activity pattern.
Three groups of market makers provide liquidity to the market.
Three groups of market takers affect market memory and are responsible for larger fluctuations.
Four groups of fundamentalists make sure that the price will not diverge too much from the fundamental value.
Four groups of chartists are further divided according to their strategy (trend following, and mean reversion).
Finally, the largest group consists of noise traders.

The realizations of the agent-based model are saved in a LOBSTER-style format, with messages and book snapshots.
The main difference from the LOBSTER format is that we do not record executions, but instead record all the market orders, as well as aggressive limit orders.
It is because we predict investor actions, and not limit order events.
In the end, we only have three types of messages: (i) market orders, (ii) limit orders, and (iii) cancellations.
Each message in \cite{nagy2023generative} contained 9 fields: \textit{type}, \textit{direction}, \textit{price}, \textit{size}, \textit{time difference}, \textit{time}, and three reference fields (price, size, and time of the referred message).
We add an extra field, called \textit{target}, which specifies whether a given action was performed by the imitated investor.
As a result, we have 23 tokens: price is described by two tokens (side and price difference), time is described 5 tokens (2 for seconds, 3 for nanoseconds, time difference is described with 4 tokens (1 for seconds, 3 for nanoseconds, and the rest is described by a single token each.
Book snapshots are stored in a sparse representation of $M$ volume features around the mid-price (we only observe $M$ depth of the order book).
Additionally, each book snapshot is paired with a mid-price change, between current and previous snapshot.

Due to token based encoding of the messages, we truncate both price changes and volumes.
Price differences that go above 999 (below -999) are set to 999 (-999).
Volumes that go above 999 shares are set to 999.
Both cases are extremely rare in our simulations and affect less than 0.01\% of actions.

\subsection{Generative model}

The literature on generative models for order-driven markets is growing \cite{hultin2023generative, nagy2023generative, wheeler2024marketgpt}.
As an initial step, we decided to build on the architecture proposed in \cite{nagy2023generative}, and further extended in \cite{nagy2025lob}.
In our case, instead of predicting the next message in the limit order book, we predict the next message of the targeted agent.
The main building block is the simplified structured state-space layer (S5) \cite{smith2022simplified}, which is based on state-space modeling \cite{durbin2012time}, known from control theory.
The reason for using state-space models, instead of widely accepted transformers \cite{vaswani2017attention}, is that the latter has quadratic complexity in the sequence length $L$, while the former is linear in $L$.
This is particularly important when dealing with limit order book data, where enormously large sequences of data are available, and it is known that long-memory effects are present.

The input of the model consists of a sequence of messages and a sequence of order book snapshots, both of which are described in detail in Section \ref{sec:data}.
During the training, we only used message sequences, which end with an action from the agent to be imitated.
The sequence of messages is flattened and so its length is equal to the product of message length (number of tokens describing a single message) and the number of messages in a single sample.
The number of snapshots is equal to the number of messages, but the snapshot describes the market state just before the corresponding message.
If, for example, the last message describes order cancellation, the corresponding snapshot still contains the to-be-deleted order.

The model outputs the vector of logits with length equal to the length of the vocabulary.
In other words, model predicts the distribution over the vocabulary.
In the training procedure, we randomly choose one of the tokens in the last message, excluding time tokens and the target agent token, and mask it.
All the following tokens are hidden, so that they cannot be used in prediction.
The output is therefore the predicted distribution of the masked token.
In order to find the optimal parameters, the model is trained by minimizing cross-entropy between the mentioned distribution and the true labels.
Optimization is performed using gradient descent with the Adam optimizer \cite{kingma2014adam}.

\begin{figure}
  \centering
  \includegraphics[width=0.99\linewidth]{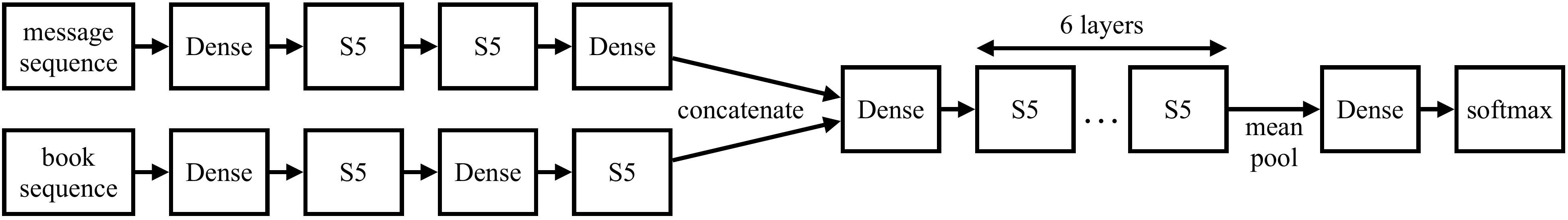}
  \caption{The architecture of the model.}
  \label{fig:architect}
\end{figure}

Following the approach from \cite{nagy2023generative}, there are two separate input branches, one for messages and one for book snapshots.
The results of the two branches are concatenated and further processed, according to the architecture shown in Fig. \ref{fig:architect}.
The training is performed for 50 epochs, using 500 messages and 500 book snapshots.
Both messages and books are projected to 128 dimensions.
The dimension of the latent space, for each used S5 layer, is set to 256.
As a result, the model contains almost $6 \cdot 10^6$ parameters.

\section{Experimental Results}

A significant plus of the multi-token messages is that we can compute distributions over different action aspects, such as price, direction, type etc.
On the other hand, some actions of the agents in simulations, have straightforward distribution.
For this reason, we mostly focus on noise traders, for which we can easily compute the distribution of price, size, direction, and action type.
While this task may seem simple, because noise traders act in a random manner, independent of the market, the model is not aware of that, and tries to find a pattern in their behavior.
For comparison, we also train the model on two types of chartists, momentum traders and mean reversion traders.
While it is more difficult to compute strict analytical distributions for them, we at least have some expectations about the values.

\begin{figure}
  \centering
  \includegraphics[width=0.49\linewidth]{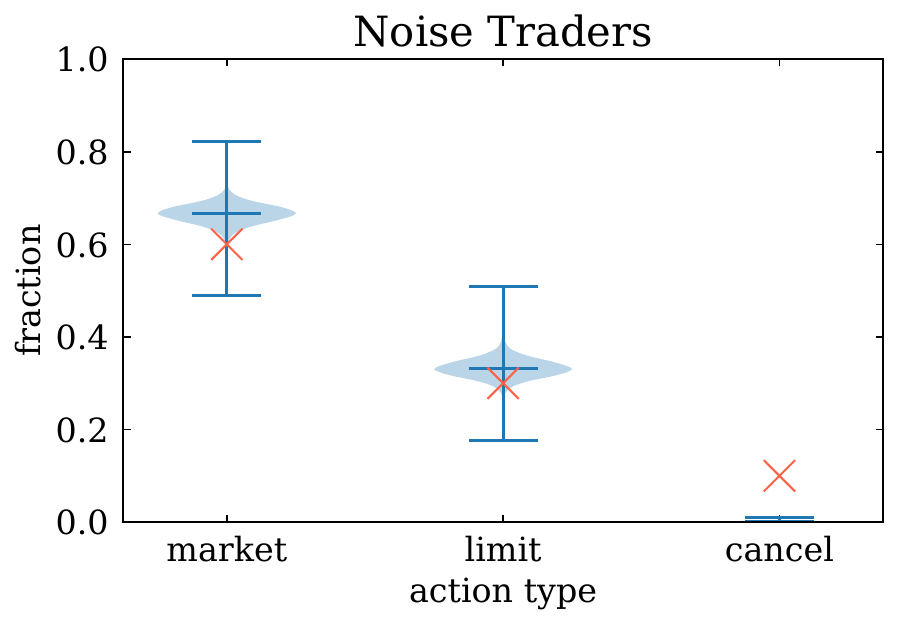}
  \includegraphics[width=0.49\linewidth]{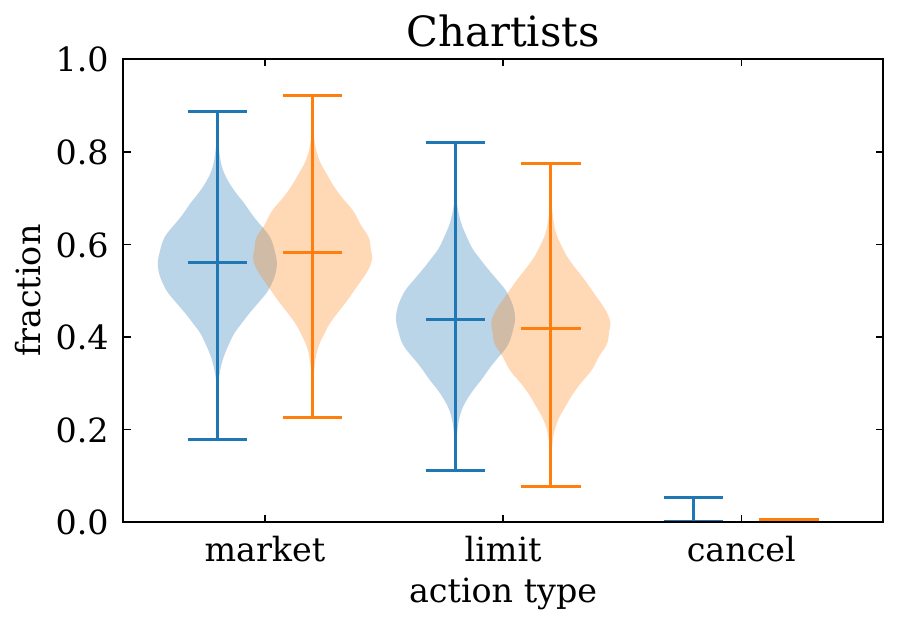}
  \caption{Fractions of different action types, computed through averaging the distributions of the model, obtained with the test data.
  \textit{Left:} Results for the model trained on noise traders with red crosses being the ground truth according to agents parametrization.
  \textit{Right:} Results obtained for two different models trained on chartists, one on momentum traders, the other on mean reversion traders.}
  \label{fig:type}
\end{figure}

First, we train three models on 1000 days of artificial trading dynamics.
Next we generate 200 additional days, which are used as test data.
For each sample (sequence of messages and book snapshots) from the test data, we compute the probabilities of each outcome (price difference, order size etc.).
Then we use these probabilities to generate violin plots for each outcome based on all the test samples.

\begin{figure}
  \centering
  \includegraphics[width=0.49\linewidth]{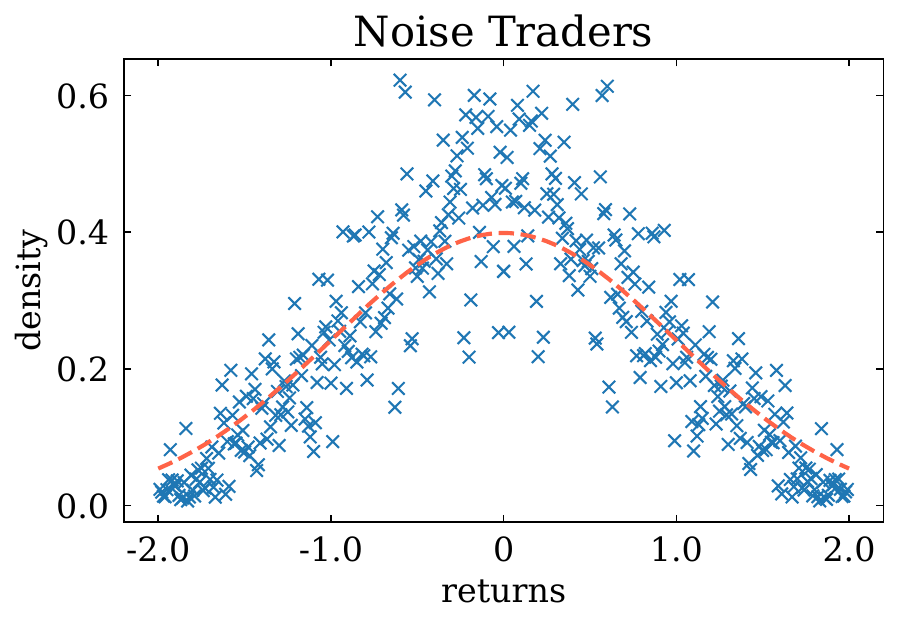}
  \includegraphics[width=0.49\linewidth]{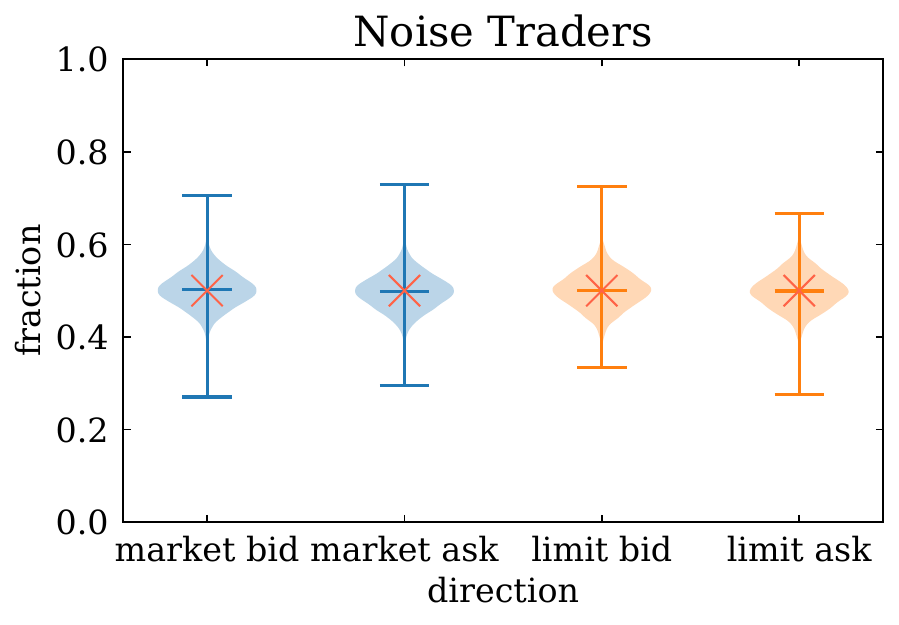}
  \caption{
  \textit{Left:} Distribution of returns for noise trader model, averaged over test data sample. Red dashed line represents the true distribution.
  \textit{Right:} Fractions of different order directions (bod or ask) for both market and limit orders. Red crosses are the ground truth according to agents parametrization.
  }
  \label{fig:ret_dir}
\end{figure}

Fig. \ref{fig:type} shows the expected fractions of different action types for both the noise trader and chartists.
In the case of noise trader model, the numbers for limit and market orders are not far from true values (marked with red crosses), but they are quite volatile.
The expected fraction of cancellation is significantly underestimated.
While we cannot precisely point to the correct values for chartists, we at least expect them to have similar values, and that is why we plot them next to each other.

\begin{figure}
  \centering
  \includegraphics[width=0.32\linewidth]{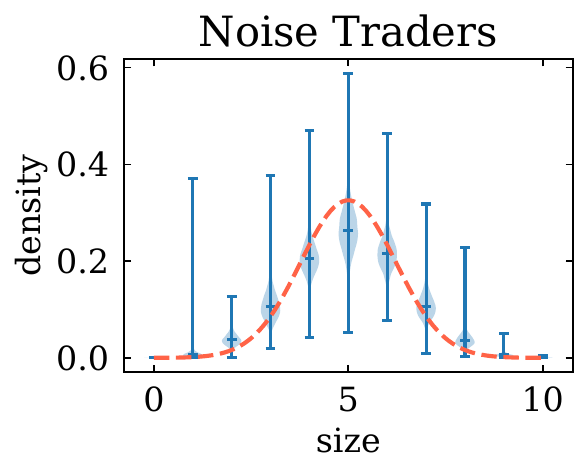}
  \includegraphics[width=0.32\linewidth]{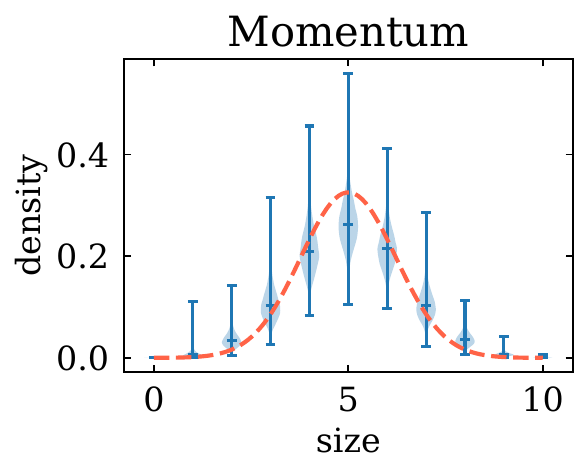}
  \includegraphics[width=0.32\linewidth]{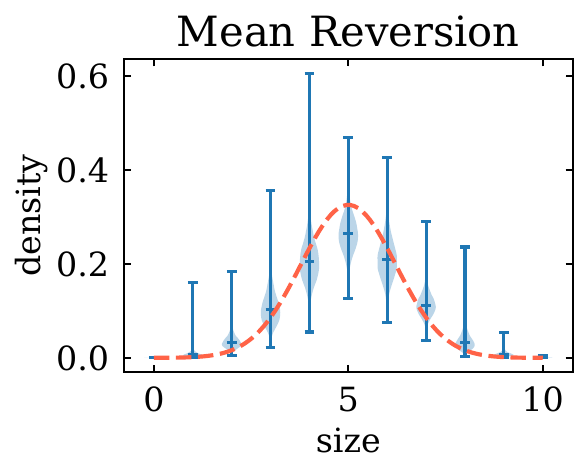}
  \caption{Distribution of order size for noise traders and chartists (momentum and mean reversion traders), averaged over test data sample.
  Red dashed line represents the true distribution.}
  \label{fig:size}
\end{figure}

Both price returns and order directions distributions are shown in Fig. \ref{fig:ret_dir}.
For price returns we only present medians in order to make the plot more readable.
The resulting distribution is narrower than the true one, but they are relatively close.
It is worth noting that the resulting distribution is very symmetric, even for values which are far from the ground truth.
The directions are well matched, but the results have broad distribution.

Distributions of order size are shown in Fig. \ref{fig:size}.
In all three cases the results nicely align with the true distributions.
Although, as before, high variability is observed.

\section{Discussions}

We present here an initial work on the prospects of using generative models for the task of imitating investors.
It does not come as a surprise that we observe high variability in estimated probabilities.
Even with only 500 messages, the number of possible configurations far exceeds the number of data points, which is a problem for estimating conditional probabilities.
Further investigation with much larger training datasets is required.

On the other hand, results mostly align with expected values.
A curious next step would be to focus more on the evaluation of conditional predictions.
This would allow for a better comparison with more complex agents, such as partly introduced chartists.

Important part, which was deliberately omitted at this stage of work, is the time dimension.
While some generative models try to predict waiting times for events in the limit order book, it is much more complex for individual agents.
At the same time, future work will need to focus on this direction, because it is of significant importance if one wants to use such trained models with agent-based environments.

\section*{Acknowledgments}

M. Wilinski acknowledges support from the European Union’s Horizon Europe programme under the Marie Skłodowska-Curie Actions (Grant Agreement No. 101066936, Project: DataABM).

\bibliographystyle{unsrt}
\bibliography{refs}

\end{document}